\documentstyle[namedreferences,proceedings]{crckapb}

\newcommand{\Rs}{r_{\rm s}}
\newcommand{\Lc}{L_{\rm c}}
\newcommand{\bfrac}[2]{\frac{\displaystyle #1}{\displaystyle#2}}
\newcommand{\kappas}{\kappa_*}

\newcommand{\CITEX}[1]{\citeauthor{#1} (\citeyear{#1})}

\begin{opening}
\title{Simulation of microlensing lightcurves \protect\\
  by combining contouring and rayshooting}
\author{Stein Vidar Hagfors Haugan}
\institute{Institute of Theoretical Astrophysics, University of Oslo\\
  Pb. 1029, Blindern\\
  N-0315 OSLO\\
  {\tt http://www.uio.no/\~\/steinhh/index.html}}
\end{opening}

\begin{document}

\section{Introduction}
The contouring methods described by
\CITEX{Lewis-MiraldaEscude-Rich-Wambs93} and
\CITEX{Witt93-contouring-APJ} are very efficient for obtaining the
magnification of a point source moving along a straight track in the
source plane. For finite sources, however, the amplification {\em
  must} be computed for numerous parallel tracks and then convolved
with the source profile.  Rayshooting, on the other hand, is an
efficient algorithm for relatively large sources, but the computing
time increases with the inverse of the source area for a given noise
level.

\section{The hybrid method}
\label{sec:method}
By using the method described in
\CITEX{Lewis-MiraldaEscude-Rich-Wambs93}, all the images of a
straight, infinite line in the source plane can be found. The images
are the borders between those parts of the lens plane projected above
the straight line, and those parts projected below the straight line.
After finding the images of one line below the source and one line
above the source, it is clear that those parts of the lens plane that
are projected between the two infinite lines in the source plane are
the areas between the images of the infinite lines.

Furthermore, those segments corresponding to the upper and lower edges
of a box surrounding the source may be identified. The end points of
these segments are projected onto the corners of the ``source box''.
Starting from the corner points, the contouring method can be ``turned
around'' 90 degrees, and all the lines joining all the corner points
of the ``source box'' are found.  After this step, all the images of
the source box are placed within known, closed polygons.  Rayshooting
is then performed within all the closed polygons, and the lightcurve
is produced in the usual way.

\section{Efficiency}
\label{sec:efficiency}

The efficiency of the rayshooting part of the method compared to
crude, non-optimized rayshooting can be found by comparing the size of
the areas where rayshooting has to be performed.  A target area in the
source plane with length $2l$, and height $2\Rs$ gives an effective
lightcurve length $\Lc = 2l-2\Rs$, where $\Rs$ is the source radius.
The theoretical efficiency $f$ can be shown to be given by
\begin{eqnarray}
  f &\approx& \left\{
  \begin{array}[c]{l}
    (1 + \bfrac{10\sqrt{\kappas}}{\Rs} + %
    {}                        \bfrac{100\kappas}{l\Rs}) \\[.5cm]
    (1+ \bfrac{20\sqrt{\kappas}}{\Rs} + \bfrac{100\kappas}{\Rs^2})
  \end{array} \right.
  \begin{array}[l]{l}
    \mbox{For $l \gg \Rs$} \\[.5cm]
    \mbox{For $l = \Rs$, $\Lc=0$.}
  \end{array}
  \label{eq:ratio}
\end{eqnarray}

\section{Discussion}
\label{sec:discussion}

The above arguments give a theoretical efficiency factor on the order
of $10^5$ for e.g. a snapshot of the source with $\Rs=0.01$, $l=\Rs$
and $\kappas=0.4$. However, the most time-consuming task for the
hybrid method is going to be the contouring itself. For a snapshot
like the example above, the contouring amounts to about $10^5$ shots
\cite{Lewis-MiraldaEscude-Rich-Wambs93}. This must be compared with
the total number of shots necessary to get a specific signal to noise
ratio, generally about $10^3$ shots. The highest estimates of $f$ thus
have to be lowered by roughly a factor of $100$, depending on the
specific parameters $\Rs$, $\kappas$, $\gamma$, and $l$.

Even so, the proposed hybrid method has the potential to be a very
efficient workhorse for producing accurate model lightcurves for small
but extended sources.

\section*{Acknowledgments}
The author would like to thank Rolf Stabell and Sjur Refsdal for
comments during the preparation of this poster.


\begin{thebibliography}{}

\bibitem[\protect\citeauthoryear{Lewis
  et~al.}{1993}]{Lewis-MiraldaEscude-Rich-Wambs93}
Lewis, G.~F., Miralda-Escude, J., Richardson, D.~C., Wambsganss, J. 1993,
  MNRAS, 261, 647

\bibitem[\protect\citeauthoryear{Witt}{1993}]{Witt93-contouring-APJ}
Witt, H.~J. 1993, ApJ, 403, 530

\end{thebibliography}
\end{document}